\documentclass[12pt]{article}
\usepackage{eurosym}
\usepackage[left=0.7in, right=0.7in, top=0.7in, bottom=0.7in]{geometry}
\usepackage{amssymb,amsmath,amsthm,color,graphicx,graphicx,cite}
\usepackage[caption=false]{subfig}
\usepackage{hyperref}
\usepackage{enumerate}
\usepackage{graphicx}
\usepackage{dsfont}
\usepackage{times}
\usepackage{bbold}
\usepackage{mathtools}
\usepackage{graphicx}
\usepackage{mathrsfs}
\usepackage{graphicx}
\usepackage{caption}
\usepackage{epstopdf}
\usepackage{epsfig}
\usepackage{eurosym}

\begin{document}

	\begin{center}
		\textbf{\Large Non-Markovian Protection and Thermal Fragility of Quantum Resources in a Spin-$1/2$ Ising–Heisenberg Diamond Chain}
		
		
		\vspace{1cm}
		
		Fadwa Benabdallah$^{a}${\footnote{%
				email: \textsf{\textbf{fadwa$_{-}$benabdallah@um5.ac.ma}}}},
	M. Y. Abd-Rabbou$^{b,c}${ \footnote{email: \textsf{\textbf{m.elmalky@azhar.edu.eg}}}},
	Mohammed Daoud$^{d,e}${\footnote{%
				email: \textsf{\textbf{m$_{-}$daoud@hotmail.com}}}} \\[0pt]
		\vspace{0.5cm} $^{a}$\textit{LPHE-Modeling and Simulation, Faculty of
			Sciences, Mohammed V University in Rabat, Morocco}\\[1em]
		$^{b}$School of Physics, University of Chinese Academy of Science, Yuquan Road 19A, Beijing, 100049, China.\\[1em]
		$^{c}$ Mathematics Department, Faculty of Science, Al-Azhar University, Nassr City 11884, Cairo, Egypt.\\[1em]
		$^{d}$\textit{ Department of Physics, Faculty of Sciences, University Ibn
			Tofail, K\'{e}nitra, Morocco}\\[1em]
		$^{e}$\textit{Abdus Salam International Centre for Theoretical Physics,
			Miramare, Trieste, Italy}
			
	\end{center}
	
	\baselineskip=18pt \medskip
	\begin{abstract}
		
		This research investigates the dynamics of entanglement and uncertainty-induced nonlocality in a spin-½ Ising–Heisenberg diamond chain subjected to local non-Markovian decoherence channels. By examining amplitude damping and random telegraph noise in both zero and finite temperature regimes, the study reveals nuanced distinctions in the degradation and revival of quantum correlations. The interplay between intrinsic spin couplings, thermal effects, and memory-induced coherence backflow highlights the complex behavior of quantum resources under realistic noise conditions. Concurrence emerges as a sensitive marker of entanglement recovery in dephasing environments, while uncertainty-induced nonlocality proves more resilient in high-temperature or dissipative regimes. The analysis further demonstrates that moderate thermal activation and external magnetic fields can nontrivially enhance or suppress quantum features depending on system parameters. These findings offer a detailed perspective on the robustness and complementarity of different quantum correlation measures, providing guiding principles for the design of thermally stable and noise-resilient quantum information protocols.
	\end{abstract}

	\section{Introduction}
Resource theories within quantum information science~\cite{S1,S2,Y1} establish a rigorous conceptual framework for the characterization and quantification of non-classical attributes inherent in quantum systems. Within this theoretical landscape, quantum entanglement and quantum coherence have emerged as prominent operational resources, instrumental in facilitating a diverse array of protocols across quantum communication, computation, and metrology~\cite{Horodecki2009,Nielsen2000}. Entanglement, characterized by its manifestation of nonlocal correlations~\cite{Datta2005,Datta2007}, proves indispensable for applications such as quantum teleportation and secure communication protocols~\cite{Bennett1993,Huang2011,Fouokeng2020, abd22}. Concurrently, quantum coherence, originating from the superposition principle, constitutes the foundational basis for interference phenomena pivotal to the advancement of quantum-enhanced technologies~\cite{S2}. A significant impediment to the practical implementation of quantum protocols stems from the deleterious effects of environmental noise and decoherence, which precipitate the degradation of these crucial quantum resources~\cite{Z2003}.

Low-dimensional spin chain systems serve as exemplary model platforms for the investigation of quantum correlation dynamics within physically realizable condensed matter environments~\cite{BEN2,BEN4,BEN5,BEN6,BEN7}. Heisenberg-type spin models, specifically, have garnered considerable attention attributable to their intricate phase structures and their amenability to exact analytical solutions under certain parametric regimes~\cite{HIDA94,BEN1,BEN3,BEN8}. Prominent among these are hybrid Ising-Heisenberg systems, which are characterized by the coupling of quantum Heisenberg dimers to classical Ising spins arrayed on lattices subject to specific geometric constraints~\cite{C1,J1,R1,R2,CH1,T1,T2,CA1,R3,R4,C2,F1,Z1}. Such hybrid architectures present a compelling confluence of physical pertinence and mathematical tractability, facilitating the derivation of exact solutions through established methodologies, including the transfer matrix technique~\cite{Baxter1982} and the generalized decoration-iteration transformation~\cite{Fisher1959,Syozi1972}. The Ising-Heisenberg diamond chain, in particular, constitutes a prototypical system for the theoretical exploration of thermal entanglement and quantum coherence within interacting spin networks.
	
From an experimental standpoint, multispin interactions, particularly those involving four-spin exchange terms, are of considerable consequence in the formulation of effective Hamiltonians for experimentally relevant material systems. Exemplary systems encompass adsorbed $^3$He layers on graphite substrates, hydrogen-bonded ferroelectric materials, squaric acid crystals, and specific copper oxide spin ladder compounds implicated in the phenomenon of high-temperature superconductivity~\cite{Roger1983,Honda1997,Reimers1989,Matsuda1999,Hiroi1991}. The incorporation of such higher-order couplings engenders novel pathways towards quantum frustration and facilitates the emergence of phase transitions augmented by correlations; however, the theoretical analysis of such couplings is frequently rendered intractable owing to the non-commuting nature of quantum spin operators. Conversely, the ubiquitous interaction of quantum systems with their surrounding environments inevitably induces decoherence, a process that typically leads to the suppression of non-classical correlations and thereby constrains the operational efficacy of quantum technologies~\cite{Breuer2002,Rivas2014}. Non-Markovian dynamics, distinguished by environmental interactions exhibiting temporal correlations and a retro-causal information flow from the environment to the system, are increasingly recognized as a propitious strategy for ameliorating these deleterious impacts. Notably, the mnemonic characteristics intrinsic to non-Markovian noise environments can facilitate the transient restoration of coherence and the revival of entanglement~\cite{Laine2010,Bylicka2014}. 
	
The critical imperative to preserve quantum correlations against environmental decoherence, particularly under the complex influence of non-Markovian dynamics, underpins substantial research efforts aimed at advancing quantum technologies. This study is motivated by the need for a systematic elucidation of quantum correlation dynamics within a spin-$\frac{1}{2}$ Ising-Heisenberg diamond chain~\cite{G12} when subjected to local non-Markovian decoherence. This model system is chosen for its capacity to offer both analytical tractability and physical relevance. Utilizing concurrence as a quantifier for entanglement and UIN for broader quantum correlations, this work undertakes a comparative analysis of the effects induced by two representative noise paradigms, AD and RTN. By employing the Kraus operator formalism~\cite{K71,CH75}, we meticulously derive the time- and temperature-dependent reduced density matrix of the constituent Heisenberg dimer. This derivation facilitates an in-depth exploration of the modulatory influences of key system parameters, including the Heisenberg exchange coupling, the four-spin interaction term, and externally applied magnetic fields. A primary objective is to investigate the potential for non-Markovian memory effects to engender revivals in quantum correlations, while concurrently assessing the differential efficacy of RTN versus AD in maintaining quantum coherence. Furthermore, this research aims to establish the utility of UIN as a robust indicator of non-classical behavior, particularly in regimes characterized by high thermal agitation or strong dissipative coupling, where entanglement may be entirely suppressed. An additional line of inquiry pursues the identification of conditions wherein moderate thermal excitation or the application of magnetic fields can beneficially augment quantum correlations through mechanisms such as energy level reconfiguration or partial thermal activation. The pivotal role of internal coupling ratios, specifically $J_H/J_I$ and $K/J_I$, in conferring stability to entanglement against decoherent processes is also a focal point. The anticipated findings are expected to provide crucial insights into the strategies for controlling and enhancing the resilience of quantum correlations within open spin systems, thereby offering pragmatic guidelines for maintaining coherence in solid-state quantum technological applications. 
	
The manuscript is structured as follows to delineate this investigation comprehensively. Section~\ref{Sec2} introduces the theoretical model of the spin-$\tfrac{1}{2}$ Ising–Heisenberg diamond chain and outlines the derivation of its thermal initial state. Subsequently, Section~\ref{Sec3} is dedicated to the detailed construction of the time-dependent density matrix under the influence of local non-Markovian decoherence, with explicit consideration given to both amplitude damping and random telegraph noise channels. In Section~\ref{Sec4}, formal definitions of concurrence and UIN are presented; these metrics serve as the principal measures for quantifying the quantum correlations manifest in the system. Section~\ref{Sec5} provides an extensive discussion of the salient results concerning the dynamical behavior of entanglement and UIN. Finally, Section~\ref{Sec6} offers concluding remarks, summarizing the key contributions of this work.
	
	\section{Theoretical model and its initial state}\label{Sec2}
	
\begin{figure*}[tbp]
		\centering
		\includegraphics[scale=0.4,trim=00 00 00 00, clip]{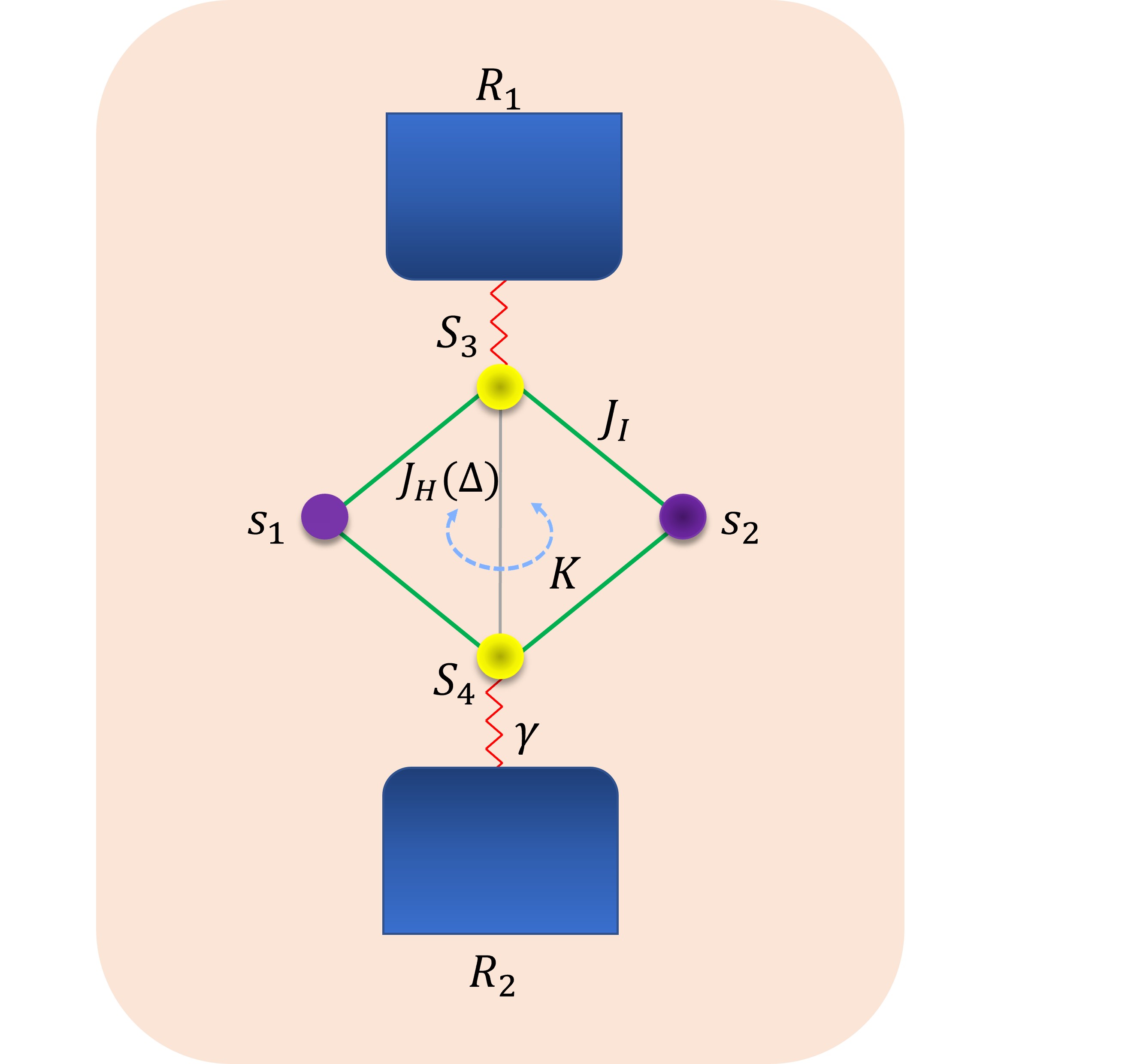} %
		\caption{A schematic representation of the structure of the spin-1/2 Ising–Heisenberg diamond chain connected to two independent baths, $R_{1}$ and $R_{2}$, at zero temperature. The yellow circles represent the Heisenberg spins, while the purple circles represent the Ising spins.}
		\label{Fig.0}
	\end{figure*}

This section delineates the Hamiltonian for a spin-$\frac{1}{2}$ Ising–Heisenberg diamond chain model, which explicitly incorporates a four-spin interaction term and is subjected to an externally applied magnetic field. The model architecture comprises Heisenberg spins ($\hat{S}_3, \hat{S}_4$) situated at interstitial positions and Ising spins ($\hat{s}_1, \hat{s}_2$) located at the nodal sites, as schematically depicted in Figure~\ref{Fig.0}. The total Hamiltonian governing the system is precisely formulated as follows~\cite{G12}
\begin{equation}\label{Hamiltonian}
	\hat{H}= J_{H}[\Delta(\hat{S}_{3}^{x}\hat{S}_{4}^{x}+\hat{S}_{3}^{y}\hat{S}_{4}^{y})+\hat{S}_{3}^{z}\hat{S}_{4}^{z}] + J_{I}(\hat{s}_{1}^{z}+\hat{s}_{2}^{z})(S_{3}^{z}+S_{4}^{z}) + K \hat{s}_{1}^{z}\hat{s}_{2}^{z}\hat{S}_{3}^{z}\hat{S}_{4}^{z} - h_{H}(\hat{S}_{3}^{z}+\hat{S}_{4}^{z}) - \frac{h_{I}}{2}(\hat{s}_{1}^{z}+\hat{s}_{2}^{z}).
\end{equation}
Here, $J_{H}$ denotes the coupling strength of the anisotropic XXZ interaction acting between nearest-neighbor Heisenberg spins; the anisotropy parameter $\Delta$ governs the specific nature of this exchange interaction. Specifically, the condition $\Delta<1$ corresponds to an easy-axis anisotropic regime, whereas $\Delta>1$ indicates an easy-plane anisotropy. The parameter $J_{I}$ quantifies the strength of the Ising-type interaction between the Heisenberg spins and their adjacent Ising spin counterparts. Furthermore, $K$ represents the coupling constant for a four-spin Ising interaction term, which involves both Heisenberg spins and the two Ising spins constituting each diamond-shaped plaquette. The external magnetic fields, $h_{I}$ and $h_{H}$, are applied to the Ising and Heisenberg spins, respectively, and are both oriented along the conventional $z$-axis. 
The Hamiltonian, when expressed in its matrix representation $\hat{\mathcal{H}}$, exhibits the following block-diagonal form:
\begin{equation}
	\hat{\mathcal{H}}=\left(
	\begin{array}{cccc}
		\hat{\mathcal{H}}_{1} & \mathfrak{0} & \mathfrak{0} &\mathfrak{0} \\
		\mathfrak{0} & \hat{\mathcal{H}}_{0} & \mathfrak{0} & \mathfrak{0} \\
		\mathfrak{0} & \mathfrak{0} & \hat{\mathcal{H}}_{0} & \mathfrak{0} \\
		\mathfrak{0} & \mathfrak{0} & \mathfrak{0} & \hat{\mathcal{H}}_{-1}%
	\end{array}
	\right).
\end{equation}
The constituent block sub-matrices within this framework are defined by distinct configurations of the nodal Ising spins ($\hat{s}_{1}^{z}, \hat{s}_{2}^{z}$). The sub-matrix $\hat{\mathcal{H}}_{1}$ corresponds to the sector wherein both nodal Ising spins are oriented parallel to the positive $z$-axis, characterized by $\hat{s}_{1}^{z}=+1/2, \hat{s}_{2}^{z}=+1/2$. The sub-matrix $\hat{\mathcal{H}}_{0}$ pertains to sectors where the nodal Ising spins adopt antiparallel orientations, characterized by $\hat{s}_{1}^{z}=\pm1/2$ and $\hat{s}_{2}^{z}=\mp1/2$. Finally, $\hat{\mathcal{H}}_{-1}$ represents the sector in which both nodal Ising spins are oriented parallel to the negative $z$-axis, characterized by $\hat{s}_{1}^{z}=-1/2, \hat{s}_{2}^{z}=-1/2$.
	
Subsequent to the diagonalization of each sub-Hamiltonian $\hat{\mathcal{H}}_{\alpha}$ (where $\alpha \in \{-1,0,1\}$ indexes the distinct configurations of the nodal Ising spins), the energy eigenvalues corresponding to the interstitial Heisenberg spin dimer subsystem are determined as follows:
\begin{align}
	\mathcal{E}_{1,4} &= \frac{J_H}{4} + \frac{K}{4} \mu \mp h_H \pm \left( J_I \mp \frac{h_I}{2} \right) \nu, \\
	\mathcal{E}_{2,3} &= -\frac{J_H}{4} - \frac{K}{4} \mu \pm \frac{\Delta J_H}{2} - \frac{h_I}{2} \nu,
	\label{Eigenvalues}
\end{align}
wherein $\mu = \hat{s}_{1}^{z} \hat{s}_{2}^{z}$ and $\nu =  \hat{s}_{1}^{z}+ \hat{s}_{2}^{z}$ are parameters. It is pertinent to observe that these energy eigenvalues, $\mathcal{E}_{i}$, are explicitly dependent on $\mu$ and $\nu$, which encapsulate the influence of the specific nodal Ising spin configurations. The parameter $\mu$ can assume values from the set $\{\pm 1/4\}$, while $\nu$ is restricted to values from $\{0, \pm 1\}$. A value of $\nu=0$ corresponds to an antiparallel alignment of the nodal Ising spins ($s_1^z = -s_2^z$), whereas $\nu=\pm1$ signifies a parallel alignment thereof ($ \hat{s}_1^z =  \hat{s}_2^z = \pm 1/2$). Employing the standard notation where $|\uparrow\rangle$ and $|\downarrow\rangle$ represent individual spin-up and spin-down states, respectively (relative to the $z$-axis), the corresponding orthogonal eigenstates of the Heisenberg spin dimer Hamiltonian within each $\hat{\mathcal{H}}_\alpha$ block, expressed in the direct product basis formed by these single-spin states (often referred to as the standard or computational basis), are given by:
\begin{equation}
	\left\vert \psi_{1}\right\rangle =\left\vert \uparrow\uparrow\right\rangle ,\text{ \ \ \ }
	\left\vert \psi_{2,3}\right\rangle =\frac{1}{\sqrt{2}}
	\left( \left\vert \uparrow\downarrow\right\rangle \pm\left\vert \downarrow\uparrow\right\rangle
	\right) ,\text{ \ \ }\left\vert \psi _{4}\right\rangle =\left\vert
	\downarrow\downarrow
	\right\rangle.  \label{Eigenvectores}
\end{equation}

	At thermal equilibrium, the initial state of the spin-1/2 Ising–Heisenberg diamond chain described by Hamiltonian ($\ref{Hamiltonian}$) is described by the Gibbs density operator in thermal equilibrium:
	\begin{equation}
		\hat{\rho}_{\alpha}(0, T) = \frac{1}{\mathcal{Z}} \sum_{i=1}^{4} \exp(-\beta \mathcal{E}_i) |\psi_i\rangle \langle \psi_i|,
	\end{equation}
	where $\mathcal{Z}=\texttt{Tr}[\exp(-\beta \mathcal{H})]$ is the partition function and $\beta = 1/(k_B T)$ is the inverse temperature, with $k_B$  is the Boltzmann constant. Therfore, the non-zero elements of the initial density matrix $\hat{\rho}_{\alpha}(0, T)$, written in the standard basis $ \{ |00\rangle, |01\rangle, |10\rangle, |11\rangle \} $, are computed as follows:
	\begin{align}
		\rho^{0}_{1,1} &= \frac{1}{\mathcal{Z}} \exp(-\beta \mathcal{E}_{1}), \\
		\rho^{0}_{2,2} &= \frac{1}{2\mathcal{Z}} \left(\exp(-\beta \mathcal{E}_{2}) + \exp(-\beta \mathcal{E}_{3})\right), \quad \rho^{0}_{3,3} = \rho^{0}_{2,2}, \\
		\rho^{0}_{2,3} &= \frac{1}{2\mathcal{Z}} \left( \exp(-\beta \mathcal{E}_{2}) - \exp(-\beta \mathcal{E}_{3}) \right), \quad \rho^{0}_{3,2} = \rho^{0^*}_{2,3}, \\
		\rho^{0}_{4,4} &= \frac{1}{\mathcal{Z}} \exp(-\beta \mathcal{E}_{4}).
	\end{align}
	Here, $\rho^{0}_{1,1}$, $\rho^{0}_{2,2}$, and $\rho^{0}_{4,4}$ correspond to the populations of the states $ |\psi_1\rangle = |00\rangle $, $ |\psi_{2,3}\rangle = (|01\rangle \pm |10\rangle)/\sqrt{2} $, and $|\psi_4\rangle = |11\rangle $, respectively. The element $\rho^{0}_{2,3}$ represents the coherence between the degenerate Bell-like states. 
	\section{Time-dependent density matrix}
	\label{Sec3}
To investigate the dynamics of quantum correlations under thermal influence, we consider a model in which the Heisenberg spins, $S_3$ and $S_4$, interact independently with their respective local environments. The temporal evolution of this open quantum system is described using the operator-sum, or Kraus, representation —a formalism suitable for characterising non-unitary dynamics arising from decoherent processes~\cite{K71,CH75}. Within this framework, the density matrix of the two-qubit Heisenberg subsystem at time $t$ and temperature $T$, denoted $\rho_{\alpha}(t,T)$, is determined from its initial state $\rho_{\alpha}(0,T)$ via the transformation:
\begin{equation}
	\rho_{\alpha}(t,T)=\sum^{1}_{i=0}\sum^{1}_{j=0}(\mathcal{K}_{i}(t,0)\otimes \mathcal{K}_{j}(t,0))\rho_{\alpha}(0,T)(\mathcal{K}_{i}(t,0)\otimes \mathcal{K}_{j}(t,0))^{\dag},  \label{TTDD}
\end{equation}
where $\mathcal{K}_{i}(t,0)$ and $\mathcal{K}_{j}(t,0)$ are the Kraus operators corresponding to the local noise channel acting independently on each constituent spin. These operators are drawn from a set $\{\mathcal{K}_k(t,0)\}_{k=0,1}$ that fulfills the trace-preservation condition $\sum_{k=0}^{1}\mathcal{K}_{k}(t,0)\mathcal{K}^{\dag}_{k}(t,0)=I$, with $I$ representing the identity operator for a single qubit.
	
In this work, we consider two paradigmatic quantum channels:\\
(i) The AD channel, selected for its representation of dissipative dynamics involving energy exchange with, and typically irreversible energy loss to, the surrounding thermal reservoir.\\
(ii) The RTN channel, chosen to model pure dephasing phenomena that arise from classically stochastic fluctuations in the local environment experienced by the qubits.\\
Crucially, the theoretical treatment of both these channels is undertaken within the non-Markovian regime. This approach is adopted to explicitly incorporate memory effects inherent in the system-environment interaction, thereby permitting the potential for information retrodiction, or backflow, from the environment to the quantum system.
	
\subsection{Amplitude Damping (AD) Channel}
The AD channel models energy dissipation phenomena, such as spontaneous emission, involving the irreversible transfer of energy from the quantum system to its surrounding reservoir. Its non-Markovian formulation is defined by the subsequent set of Kraus operators~\cite{77}
\begin{equation}
	\mathcal{K}_0^{\mathrm{AD}}(t,0) =
	\begin{pmatrix}
		1 & 0 \\
		0 & \sqrt{1 - \lambda(t)}
	\end{pmatrix}, \quad
	\mathcal{K}_1^{\mathrm{AD}}(t,0) =
	\begin{pmatrix}
		0 & \sqrt{\lambda(t)} \\
		0 & 0
	\end{pmatrix},
\end{equation}
where the time-dependent decoherence function $\lambda(t)$ is given by
\begin{equation}
	\lambda(t) = 1 - e^{-g t} \left[ \cosh\left(\frac{lt}{2}\right) + \frac{g}{l} \sinh\left(\frac{lt}{2}\right) \right]^2, \quad
	l = \sqrt{g(g - 2\gamma)}.
\end{equation}
Herein, $\gamma$ denotes the coupling strength between the system and the reservoir, intrinsically linked to the system's relaxation timescale $\tau_s = 1/\gamma$. The parameter $g$ represents the spectral width of the reservoir, which is inversely proportional to the reservoir's correlation time, $\tau_r = 1/g$. The dynamical characteristics of the system under the influence of amplitude damping are determined by the relative magnitudes of the coupling strength $\gamma$ and the reservoir's spectral bandwidth $g$. In scenarios where the system-reservoir coupling is substantially weaker than the spectral width of the environment (i.e., $2\gamma \ll g$), the reservoir effectively possesses a short memory, and consequently, the system dynamics approximate a Markovian process. Conversely, under conditions where $2\gamma \gg g$, memory effects within the reservoir become significant, engendering non-Markovian dynamics. These dynamics are notably characterized by the potential for transient information retrodiction from the environment back to the quantum system~\cite{78}.

Subject to the influence of AD noise, the two-qubit density matrix, denoted $\rho^{\mathrm{AD}}_\alpha(t,T)$ and expressed in the standard computational basis, evolves while preserving a specific block-diagonal structure:
\begin{equation}
	\rho^{\mathrm{AD}}_\alpha(t,T) =
	\begin{pmatrix}
		\rho_{1,1}^{\mathrm{AD}} & 0 & 0 & 0 \\
		0 & \rho_{2,2}^{\mathrm{AD}} & \rho_{2,3}^{\mathrm{AD}} & 0 \\
		0 & \rho_{3,2}^{\mathrm{AD}} & \rho_{3,3}^{\mathrm{AD}} & 0 \\
		0 & 0 & 0 & \rho_{4,4}^{\mathrm{AD}}
	\end{pmatrix},
	\label{R1}
\end{equation}
with the explicit expressions for the time-dependent matrix elements given by:
\begin{align}
	\rho_{1,1}^{\mathrm{AD}} &= \rho^{0}_{1,1} + 2\rho^{0}_{2,2} \lambda(t) + \rho^{0}_{4,4} \lambda(t)^2, \\
	\rho_{2,2}^{\mathrm{AD}} &= \rho^{0}_{2,2} \left(1 - \lambda(t)\right) + \rho^{0}_{4,4} \lambda(t)\left(1 - \lambda(t)\right), \\
	\rho_{2,3}^{\mathrm{AD}} &= \rho^{0}_{2,3} \left(1 - \lambda(t)\right), \quad \rho_{3,2}^{\mathrm{AD}} = (\rho_{2,3}^{\mathrm{AD}})^*, \\
	\rho_{3,3}^{\mathrm{AD}} &= \rho_{2,2}^{\mathrm{AD}}, \quad \rho_{4,4}^{\mathrm{AD}} = \rho^{0}_{4,4} \left(1 - \lambda(t)\right)^2.
\end{align}
The terms $\rho^{0}_{ij}$ denote the corresponding elements of the initial density matrix $\rho_{\alpha}(0,T)$.
	
	\subsection{Random Telegraph Noise (RTN) Channel}
The random telegraph noise (RTN) channel serves to model pure dephasing processes, which are precipitated by classical, bistable stochastic fluctuations originating from the local environment. For this noise model, the Kraus operators acting on a single qubit are formulated as follows~\cite{4,38}:
\begin{equation}
	\mathcal{K}_0^{\mathrm{RTN}}(t,0) = \sqrt{\frac{1 + \Lambda(t)}{2}} \, \mathbb{I}, \quad
	\mathcal{K}_1^{\mathrm{RTN}}(t,0) = \sqrt{\frac{1 - \Lambda(t)}{2}} \, \sigma_z,
\end{equation}
where $\mathbb{I}$ is the identity operator and $\sigma_z$ is the Pauli-Z matrix. The time-dependent decoherence function $\Lambda(t)$ is specified by:
\begin{equation}
	\Lambda(t) = e^{-\gamma t} \left[ \cos(\zeta \gamma t) + \frac{\sin(\zeta \gamma t)}{\zeta} \right], \quad \text{with} \quad \zeta = \sqrt{\left(\frac{2b}{\gamma}\right)^2 - 1}.
\end{equation}
In this formulation, $b$ parameterizes the coupling strength between the qubit and the environmental fluctuator, while $\gamma$ denotes the rate at which this fluctuator undergoes stochastic transitions between its states. Within the RTN framework, the qualitative nature of the system's dynamics is dictated by the dimensionless parameter $4b\tau$, where $\tau = 1/(2\gamma)$ represents the characteristic correlation time of the stochastic noise. The system dynamics are characterized as Markovian under the condition $(4b\tau)^2 < 1$, a regime indicative of rapidly fluctuating environmental noise wherein memory effects are negligible. Conversely, if $(4b\tau)^2 > 1$, temporal correlations inherent in the noise process become substantial, thereby precipitating non-Markovian dynamics distinguished by discernible memory effects~\cite{39}.

When subjected to RTN, the density matrix, $\rho^{\mathrm{RTN}}_\alpha(t,T)$, preserves the block-diagonal structure previously described; however, the temporal evolution of its constituent elements under RTN is distinct:
\begin{equation}
	\rho^{\mathrm{RTN}}_\alpha(t,T) =
	\begin{pmatrix}
		\rho^{\mathrm{RTN}}_{1,1} & 0 & 0 & 0 \\
		0 & \rho^{\mathrm{RTN}}_{2,2} & \rho^{\mathrm{RTN}}_{2,3} & 0 \\
		0 & \rho^{\mathrm{RTN}}_{3,2} & \rho^{\mathrm{RTN}}_{3,3} & 0 \\
		0 & 0 & 0 & \rho^{\mathrm{RTN}}_{4,4}
	\end{pmatrix}.
	\label{R2}
\end{equation}
The non-zero matrix elements are given by:
\begin{align}
	\rho^{\mathrm{RTN}}_{1,1}(t,T) &= \rho^{0}_{1,1}, \\
	\rho^{\mathrm{RTN}}_{2,2}(t,T) &= \rho^{0}_{2,2}, \\
	\rho^{\mathrm{RTN}}_{3,3}(t,T) &= \rho^{\mathrm{RTN}}_{2,2}(t,T) \quad (= \rho^{0}_{2,2}), \\
	\rho^{\mathrm{RTN}}_{4,4}(t,T) &= \rho^{0}_{4,4}, \\
	\rho^{\mathrm{RTN}}_{2,3}(t,T) &= \rho^{0}_{2,3} \cdot \Lambda(t)^2, \\
	\rho^{\mathrm{RTN}}_{3,2}(t,T) &= (\rho^{\mathrm{RTN}}_{2,3}(t,T))^*.
\end{align}
Here, $\rho^{0}_{ij}$ denote the elements of the initial density matrix $\rho_{\alpha}(0,T)$. In the context of the RTN channel, the diagonal population terms $\rho^{\mathrm{RTN}}_{1,1}$, $\rho^{\mathrm{RTN}}_{2,2}$, $\rho^{\mathrm{RTN}}_{3,3}$ (which is equivalent to $\rho^{\mathrm{RTN}}_{2,2}$ and thus to $\rho^{0}_{2,2}$), and $\rho^{\mathrm{RTN}}_{4,4}$ do not exhibit temporal evolution dependent on the decoherence function $\Lambda(t)$. In contrast, the off-diagonal coherence terms, $\rho^{\mathrm{RTN}}_{2,3}$ and its complex conjugate $\rho^{\mathrm{RTN}}_{3,2}$, are attenuated over time by the factor $\Lambda(t)^2$. In the RTN case, the diagonal elements remain unaffected while the off-diagonal coherence terms decay over time, reflecting the pure dephasing nature of the channel.
	
\subsection{Total Time-Dependent Density Matrix}
	For both delineated noise paradigms (Amplitude Damping, AD, and Random Telegraph Noise, RTN), the effective thermal density matrix of the Heisenberg spin dimer ($S_3, S_4$) subsystem, embedded within the infinite spin-$\frac{1}{2}$ Ising–Heisenberg chain and subjected to local environmental interactions, is constructed. This matrix, representing an average over the distinct, equally probable classical configurations of the adjacent nodal Ising spins ($s_1, s_2$), assumes the following block-diagonal structure:
	\begin{equation}
		\rho^{X}(t,T) = \frac{1}{4}
		\begin{pmatrix}
			\rho^{X}_{+1}(t,T) & \mathfrak{0} & \mathfrak{0} & \mathfrak{0} \\
			\mathfrak{0} & \rho^{X}_{0}(t,T) & \mathfrak{0} & \mathfrak{0} \\
			\mathfrak{0} & \mathfrak{0} & \rho^{X}_{0}(t,T) & \mathfrak{0} \\
			\mathfrak{0} & \mathfrak{0} & \mathfrak{0} & \rho^{X}_{-1}(t,T)
		\end{pmatrix},
	\end{equation}
	where $X \in \{\text{AD, RTN}\}$ specifies the operative noise channel. The constituent block matrices, $\rho^{X}_{\alpha}(t,T)$ with $\alpha \in \{0, \pm1\}$, denote the time-evolved reduced density matrices of the Heisenberg spin dimer ($S_3, S_4$). Each such reduced density matrix is conditioned upon a specific, fixed classical spin configuration of the neighboring Ising spins $s_1$ and $s_2$, which collectively determine the parameter $\alpha$. Within the framework of this hybrid classical–quantum model, decoherent evolution is exclusively experienced by the quantum subsystem (the Heisenberg dimer), whereas the classical Ising spins are considered static throughout the interaction. This methodological approach facilitates a systematic analytical investigation into the complex interplay among thermal effects, quantum noise characteristics, and inherent spin interaction parameters, as they collectively influence the dynamical behavior of quantum correlations within such hybrid spin system architectures.
	
	\section{Quantum information resources}
	
	\label{Sec4}
	In this section, we explore the dynamics of quantum correlations within the spin-$\tfrac{1}{2}$ Ising–Heisenberg diamond chain in the presence of environmental decoherence. Specifically, we focus on two key quantum information resources, namely,  entanglement by means of concurrence and UIN. Our aim is to analyze how these quantities evolve under thermal fluctuations and non-Markovian noise, modeled via AD and RTN channels that act locally on the Heisenberg spins.
	
	\subsection{Quantum entaglement: Concurrence}
	To quantify bipartite entanglement, we employ the concurrence $\mathcal{C}$, a widely used  quantum information metric introduced by Wootters~\cite{W1997,W1998}. For a general two-qubit density matrix $\rho$, the concurrence is given by
	\begin{equation}
		\mathcal{C}(\rho) = \max \left\{ 0,\; \sqrt{\lambda_1} - \sqrt{\lambda_2} - \sqrt{\lambda_3} - \sqrt{\lambda_4} \right\},
	\end{equation}
	where $\lambda_i$ (with \(i = 1, 2, 3, 4\)) are the eigenvalues in decreasing order of the following non-Hermitian matrix:
	\begin{equation}
		R = \rho \, (\sigma^y \otimes \sigma^y) \, \rho^* \, (\sigma^y \otimes \sigma^y),
	\end{equation}
	with $\rho^*$ representing the complex conjugate of $\rho$ in the computational basis, and $\sigma^y$ denoting the Pauli Y matrix.
	
	In our study, the time and temperature dependencies density matrices possess an X-shaped structure ($\ref{R1}, \ref{R2}$), 
	which allows for a simplified analytical expression of the concurrence. Specifically, it reduces to
	\begin{equation}
		\mathcal{C}^{X}(\rho_{\alpha}(t,T)) = 2 \max \left\{ |\rho^{X}_{2,3}| - \sqrt{\rho^{X}_{1,1} \rho^{X}_{4,4}}, 0 \right\}.
		\label{C}
	\end{equation}
	This expression indicates that only the off-diagonal element $\rho^{X}_{2,3}$ contributes to the entanglement dynamics, since the coherence $\rho^{X}_{1,4}$ remains zero for all times due to the initial structure of the thermal state and the nature of the system's evolution.
	
	\subsection{Uncertainty-induced quantum nonlocality}
	Uncertainty-induced quantum nonlocality (UIN) is a recently introduced measure of quantum correlations~\cite{L11,W14}, grounded in the concept of Wigner--Yanase skew information (WYSI)~\cite{W63}. This measure quantifies the degree of nonlocality arising from the incompatibility between a local observable and the global quantum state. In contrast to quantum discord~\cite{O01,H01}, which requires optimization over local measurement bases, UIN avoids such minimization. This makes it computationally more tractable, particularly in systems with high symmetry or constrained dynamics~\cite{L11,W14}.
	
	Given a bipartite quantum state $\rho$ on $\mathcal{H}_A \otimes \mathcal{H}_B$, the UIN with respect to subsystem $A$ is defined as
	\begin{equation}
		\mathcal{U}_C(\rho) = \max_{K_A^C \otimes I_B} I\left(\rho, K_A^C \otimes I_B\right),
	\end{equation}
	where the maximization is taken over all Hermitian operators $K_A^C$ with nondegenerate spectra that commute with the reduced density matrix $\rho_A = \mathrm{Tr}_B(\rho)$. Here, $I(\rho, K)$ denotes the Wigner--Yanase skew information:
	\begin{equation}
		I(\rho, K) = -\frac{1}{2} \mathrm{Tr} \left[ \sqrt{\rho}, K \right]^2,
	\end{equation}
	which quantifies the quantum uncertainty associated with the observable $K$ in the state $\rho$.
	
	In our model, a spin-$\tfrac{1}{2}$ Ising--Heisenberg diamond chain subjected to environmental decoherence, the system density matrix preserves an X-state structure under both thermal equilibrium and dynamical evolution, including AD $(\ref{R1})$ and RTN $(\ref{R2})$. This structure permits a closed-form evaluation of UIN as:
	\begin{equation}
		\mathcal{U}^{X}_C(\rho_{\alpha}(t,T)) =
		\begin{cases}
			1 - \omega_{\min}(W^{X}_{AB}), & \text{if } \vec{r} = \vec{0}, \\
			1 - \frac{1}{\|\vec{r}\|^2} \vec{r}^{\,T} W^{X}_{AB} \vec{r}, & \text{if } \vec{r} \neq \vec{0},
		\end{cases}
	\end{equation}
	where $\vec{r} = [r_i]$ is the Bloch vector of subsystem $A$, defined as $r_i = \mathrm{Tr}(\rho[\sigma_i \otimes I_B])$, and $W^{X}_{AB}$ is a $3 \times 3$ real symmetric matrix with elements
	\begin{equation}
		(W^{X}_{AB})_{ij} (t,T)= \mathrm{Tr} \left[ \sqrt{\rho_{\alpha}^{X}(t,T)} \, (\sigma_i \otimes I_B) \, \sqrt{\rho_{\alpha}^{X}(t,T)} \, (\sigma_j\otimes I_B) \right],
		\label{W}
	\end{equation}
	with $\sigma_{i}$ ($i = x, y, z$) denoting the Pauli matrices acting on subsystem $A$. The quantity $\omega_{\min}(W^{X}_{AB})$ refers to the smallest eigenvalue of $W^{X}_{AB}$. Details on the explicit computation of $W^{X}_{AB}$ matrix elements are provided in Appendix A (\ref{Sec7}).
	
	The UIN measure satisfies all the essential criteria expected of a bona fide quantum correlation quantifier: it vanishes on classical-quantum states, reduces to the linear entropy for pure states, is invariant under local unitary transformations, and is non-increasing under local operations on the unmeasured subsystem~\cite{G13,A20}. Moreover, by avoiding the optimization over local measurements, UIN becomes particularly well suited for analytically tractable models such as the Ising-Heisenberg diamond chain, especially in the presence of environmental noise.

	\section{Dynamical behavior of quantum correlations under local decoherence}
	\label{Sec5}
	In this section, we analyze the time evolution of two fundamental quantum information resources: the entanglement, quantified by the concurrence $\mathcal{C}^X(t,T)$, and the nonlocality, captured via UIN $\mathcal{U}_C^X(t,T)$, within a spin-$\tfrac{1}{2}$ Ising–Heisenberg diamond chain subjected to local, non-Markovian decoherence.
	
	\begin{figure}[!h]
		\centering
		\includegraphics[scale=1.099,trim=00 00 00 00, clip]{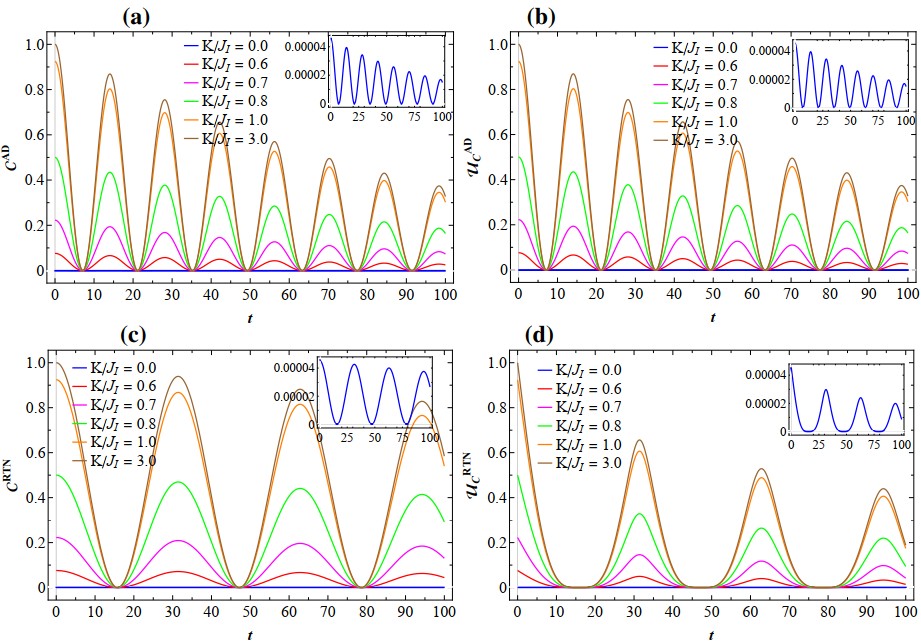} 
		\caption{ The dynamics of (a) concurrence and (b) UIN under non-Markovian AD noise for various values of $K/J_{I}$. The system parameters are: $J_{H}/J_{I} = 1.2$, $\Delta = 0.5$, $k_{B}T/J_{I} = 0.01$, $g = 0.01$, $\gamma^{\mathrm{AD}} = 10$, $h_{H}/J_{I} = 0$, and $h_{I}/J_{I} = 0$. Panels (c) and (d) present the corresponding dynamics under non-Markovian RTN with $b = 0.05$ and $\gamma^{\mathrm{RTN}} = 0.01$, respectively.}
		\label{Fig.1}
	\end{figure}
	
	Figure~\ref{Fig.1} displays the temporal dynamics of concurrence, $\mathcal{C}^{X}(t,T)$, and uncertainty-induced nonlocality, $\mathcal{U}_{C}^{X}(t,T)$, under the influence of AD channel, depicted in the upper panels, and RTN channel, illustrated in the lower panels. The temperature is fixed at a very low value, $k_B T/J_I = 0.01$, ensuring that the system initially occupies its ground state with negligible thermal excitations. At the initial time point ($t=0$), the system, under these conditions, manifests as a maximally entangled Bell-like state. Consequently, both $\mathcal{C}^{X}(0,T)$ and $\mathcal{U}_{C}^{X}(0,T)$ attain their maximal theoretical value of unity, signifying the presence of robust initial quantum correlations.  In the presence of amplitude damping (Figs.~\ref{Fig.1}(a) and~\ref{Fig.1}(b)), a channel modeling energy dissipation to the environment, both quantum correlation quantifiers exhibit a pronounced and rapid decay. Nevertheless, an augmentation of the four-spin interaction strength relative to the Ising coupling ($K/J_I$) demonstrably enhances the robustness of the quantum correlations against decoherent effects. The concurrence, $\mathcal{C}^{\mathrm{AD}}(t,T)$, displays oscillatory behavior punctuated by partial revivals, a phenomenon sometimes referred to as entanglement sudden birth (ESB). These revivals are indicative of the mnemonic characteristics inherent in the non-Markovian nature of the environmental interaction. The prominence of these revivals is further amplified by the intra-dimer Heisenberg coupling ($J_H/J_I$). This coupling term serves to increase the energy eigenlevel separation, thereby contributing to the stabilization of entangled states against deleterious decoherent influences. Following a period of transient oscillations, the concurrence typically converges towards a quasi-steady-state value, the asymptotic magnitude of which is substantially modulated by the $K/J_I$ ratio. It is noteworthy that the uncertainty-induced nonlocality, $\mathcal{U}_{C}^{\mathrm{AD}}(t,T)$, generally emulates the temporal evolution of $\mathcal{C}^{\mathrm{AD}}(t,T)$, manifesting synchronized decay profiles and revival phenomena. However, $\mathcal{U}_{C}^{\mathrm{AD}}(t,T)$ demonstrates marginally superior resilience, particularly during intervals where the concurrence transiently diminishes to zero.

	Conversely, under the influence of RTN, as depicted in Figs.~\ref{Fig.1}(c) and~\ref{Fig.1}(d)), which induces pure dephasing, the degradation of quantum resources proceeds at a comparatively attenuated rate. In this scenario, the concurrence exhibits earlier and more smoothly oscillatory revivals, a characteristic underscoring its heightened sensitivity to the initial phases of non-Markovian information backflow from the environment. Concurrently, the UIN, $\mathcal{U}_{C}^{\mathrm{RTN}}(t,T)$, undergoes a more precipitous initial decay and manifests revivals only after a discernible temporal lag. This behavior suggests a reduced responsiveness of UIN to the early stages of coherence restoration when compared to entanglement within this pure dephasing context.

	Overall, these observations suggest that RTN represents a less detrimental noise environment for the persistence of quantum correlations requisite for quantum technological applications, as it facilitates coherence preservation over more extended durations and often averts the complete eradication of entanglement. The comparative efficacy of the chosen quantum correlation quantifiers appears to be contingent upon the specific nature of the decoherence channel. Specifically, $\mathcal{U}_{C}$ demonstrates particular utility in dissipative environments, attributable to its persistence even when entanglement is fully suppressed. In contrast, $\mathcal{C}$ exhibits greater sensitivity to the recovery of quantum coherence under pure dephasing noise, rendering it a more effective indicator of the initial manifestations of non-Markovian dynamics in such scenarios.
	
	\begin{figure}[!h]
		\centering
		\includegraphics[scale=1.099,trim=00 00 00 00, clip]{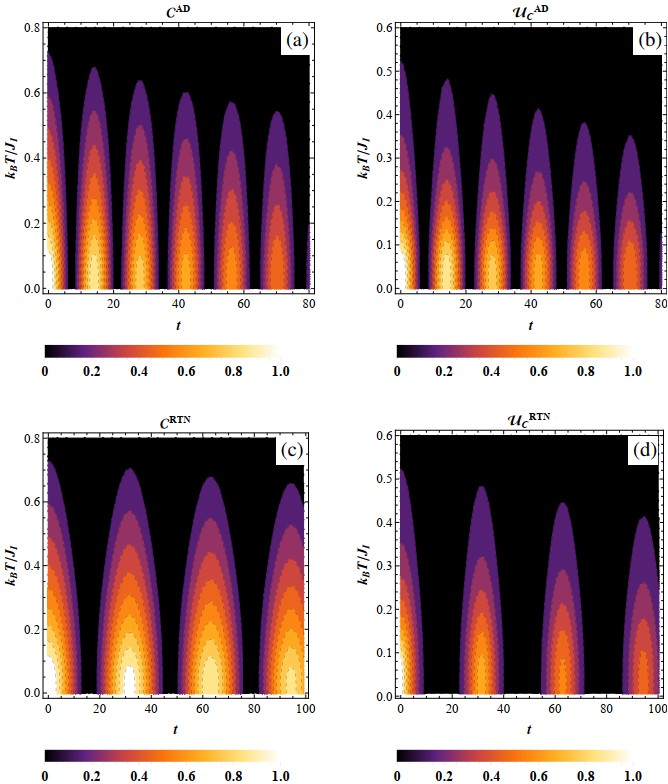} 
		\caption{The density plots of (a) concurrence and (b) UIN in the $(k_{B}T/J_{I}-t)$ plane,  under non-Markovian AD. The system parameters are: $K/J_{I} = 3.0$, $J_{H}/J_{I} = 1.2$, $\Delta = 0.5$, $g = 0.01$, $\gamma^{\mathrm{AD}} = 10$, $h_{H}/J_{I} = 0$, and $ h_{I}/J_{I} = 0$. Panels (c) and (d) present the corresponding dynamics under non-Markovian RTN with $b = 0.05$ and $\gamma^{\mathrm{RTN}} = 0.01$, respectively. }
		\label{Fig.2}
	\end{figure}
	
	Figure~\ref{Fig.2} presents a systematic mapping of the concomitant influence of thermal energy and temporal evolution on the dynamics of quantum correlations within the spin-$\tfrac{1}{2}$ Ising–Heisenberg diamond chain, subjected to local non-Markovian noise. The upper panels delineate the behavior of concurrence, $\mathcal{C}^{\mathrm{AD}}(t, T)$, and UIN, $\mathcal{U}_C^{\mathrm{AD}}(t, T)$, within the $(k_B T/J_I, t)$ parameter space for an AD channel. The lower panels provide the corresponding dynamical landscapes for $\mathcal{C}^{\mathrm{RTN}}(t, T)$ and $\mathcal{U}_C^{\mathrm{RTN}}(t, T)$ under a RTN channel. These density plots facilitate a direct quantitative comparison of the suppression and revival phenomena exhibited by quantum correlations across a continuous spectrum of temperatures and extended time scales.
	
	In the low-temperature limit ($k_B T/J_I \approx 0$), the system predominantly populates its highly entangled ground state, which corresponds to a maximally entangled Bell-type state. Consequently, both quantum correlation metrics, concurrence and UIN, are initialized at their maximum value ($\mathcal{C} = \mathcal{U}_C = 1$), indicating robust initial quantumness. With increasing temperature, thermal fluctuations induce a statistical mixture of energy eigenstates, which progressively diminishes quantum coherence and suppresses entanglement. This thermally-induced degradation is manifested across all panels as a monotonic attenuation of both measures, which are ultimately extinguished beyond a critical temperature threshold of approximately $k_B T/J_I \gtrsim 0.3$.
	
	Under the influence of amplitude damping noise (Figs.~\ref{Fig.2}(a,b)), a process characterized by irreversible energy exchange with the environment, the quantum correlations undergo a rapid decay. Notably, the concurrence, $\mathcal{C}^{\mathrm{AD}}$, experiences entanglement sudden death at approximately $k_B T/J_I \approx 0.2$, above which no subsequent revivals are observed. While $\mathcal{U}_C^{\mathrm{AD}}$ also decays swiftly, it does not exhibit superior resilience in this dissipative context; indeed, within certain temperature regimes, it vanishes even more rapidly than the concurrence. This observation is significant as it challenges the common expectation that measures of non-classicality such as UIN might persist longer than entanglement, suggesting that under dissipative conditions, entanglement serves as a particularly informative, if not more robust, resource indicator.
	
	The dynamics under the RTN channel (Figs.~\ref{Fig.2}(c,d)) are governed by pure dephasing rather than by energy dissipation. This fundamental difference results in a superior preservation of quantum coherence and gives rise to a richer landscape of dynamical features. The concurrence, $\mathcal{C}^{\mathrm{RTN}}$, displays pronounced revival patterns that persist across a broad temperature domain, particularly at low to intermediate values of $k_B T/J_I$. Intriguingly, the revival of $\mathcal{U}_C^{\mathrm{RTN}}$ is discernibly slower than that of concurrence, reflecting a diminished sensitivity to the short-timescale information backflow characteristic of the non-Markovian interaction. This temporal lag highlights a fundamental distinction between the two metrics: whereas concurrence is highly responsive to the fine-grained dynamics of quantum coherence, UIN appears to capture broader, more global aspects of system-wide correlations.
	
In general, Figure~\ref{Fig.2} underscores the critical interplay between thermal energy and the specific character of the environmental noise in dictating the longevity of quantum correlations within open spin-chain systems. Amplitude damping, being dissipative, engenders a rapid and largely irreversible loss of quantum properties. Conversely, random telegraph noise, being a dephasing mechanism, is more conducive to coherence preservation and permits partial reversibility through non-Markovian memory effects. Across both scenarios, concurrence emerges as a more sensitive and, particularly in dephasing-dominated environments, a more resilient probe of quantum correlations than uncertainty-induced nonlocality. These findings affirm the strategic value of entanglement-based diagnostics and illuminate the distinct dynamical signatures produced by different noise mechanisms, thereby providing critical insights for the design of robust quantum technologies intended for operation in thermally active environments.

	\begin{figure}[!h]
		\centering
		\includegraphics[scale=1.099,trim=00 00 00 00, clip]{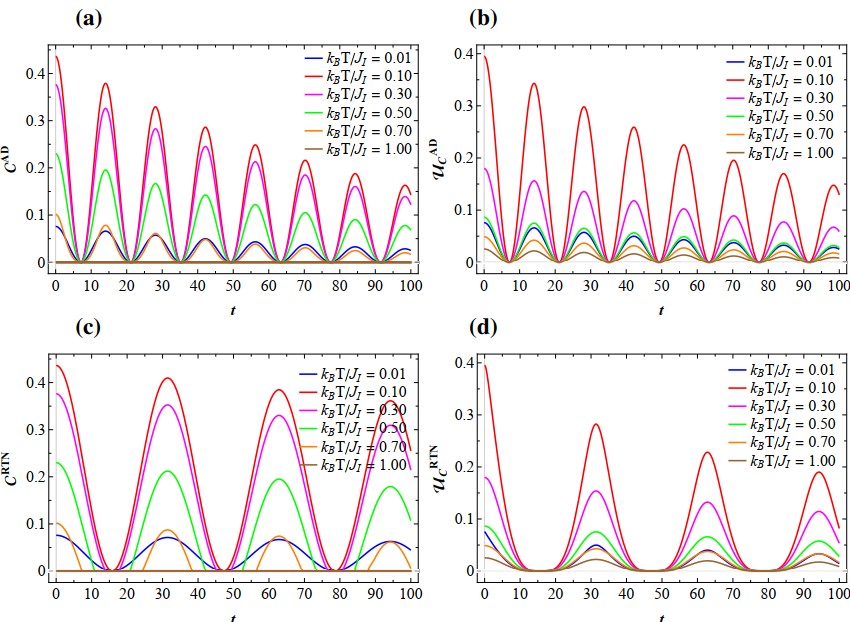} 
		\caption{ Dynamics of concurrence and uncertainty-induced non-locality under the same parameters as those used in Figure~\ref{Fig.2}, except with $h_{H}/J_{I} = 0.3$ and $h_{I}/J_{I} = 0.5$.}
		\label{Fig.3}
	\end{figure}

Figure~\ref{Fig.3} delineates the dynamical evolution of entanglement and UIN within the spin-$\tfrac{1}{2}$ Ising–Heisenberg diamond chain, now under the influence of both local non-Markovian decoherence and externally applied, axially-aligned magnetic fields. In a departure from the zero-field scenario presented in Fig.~\ref{Fig.2}, this analysis incorporates moderate field strengths, specifically $h_H/J_I = 0.3$ and $h_I/J_I = 0.5$. The introduction of these external field perturbations facilitates a targeted exploration of their impact on the resilience of quantum correlations under non-zero thermal conditions. When subjected to AD noise (Figs.~\ref{Fig.3}(a,b)), both concurrence, $\mathcal{C}^{\mathrm{AD}}(t, T)$, and UIN, $\mathcal{U}_C^{\mathrm{AD}}(t, T)$, exhibit diminished amplitudes and contracted coherence lifetimes relative to the zero-field case. Under RTN (Figs.~\ref{Fig.3}(c,d)), the dynamics remain comparatively more robust, with coherent oscillations persisting even at significant thermal energies ($k_B T/J_I \geq 0.5$). A noteworthy phenomenon is observed for both noise channels: the magnitude of quantum correlations is maximized not in the cryogenic limit, but at a finite intermediate temperature of approximately $k_B T/J_I = 0.1$. This non-monotonic temperature dependence suggests a nuanced interplay between thermal activation and the applied magnetic fields, which reconfigures the energy level structure in a manner that can transiently promote coherence. Although both correlation measures inevitably degrade with increasing temperature, the concurrence, particularly under RTN, sustains more pronounced revivals and oscillations of a larger amplitude. A particularly salient feature emerges in the high-temperature regime ($k_B T/J_I = 1$). Under these conditions, the concurrence is entirely suppressed for all time, $\mathcal{C}(t,T) \equiv 0$, signifying a complete absence of entanglement due to the combined effects of strong thermal agitation and field-induced decoherence. In stark contrast, the UIN metric, $\mathcal{U}_C(t, T)$, sustains non-zero, albeit attenuated, oscillations. This persistence of non-zero UIN in the absence of any measurable entanglement provides unambiguous evidence that certain forms of quantum correlation, distinct from entanglement, endure and remain dynamically active. This observation is valid for both decoherence channels. It establishes UIN as a more encompassing indicator of non-classicality, capable of discerning subtle quantum characteristics that are inaccessible to entanglement-based measures alone.

In summation, Figure~\ref{Fig.3} demonstrates that the decay and revival profiles of quantum correlations are intricately dependent on the specific decoherence mechanism, the thermal energy, and the applied magnetic field. Whereas both concurrence and UIN serve as potent indicators of quantum correlations at low-to-moderate temperatures, UIN uniquely retains non-trivial dynamics in the high-temperature regime where entanglement is fully extinguished. This result highlights the significant potential of UIN as a valuable diagnostic tool for identifying residual quantum correlations in physically relevant thermal environments, particularly in scenarios where conventional entanglement criteria prove insufficient.

	\begin{figure}[!h]
		\centering
	\includegraphics[scale=1.099,trim=00 00 00 00, clip]{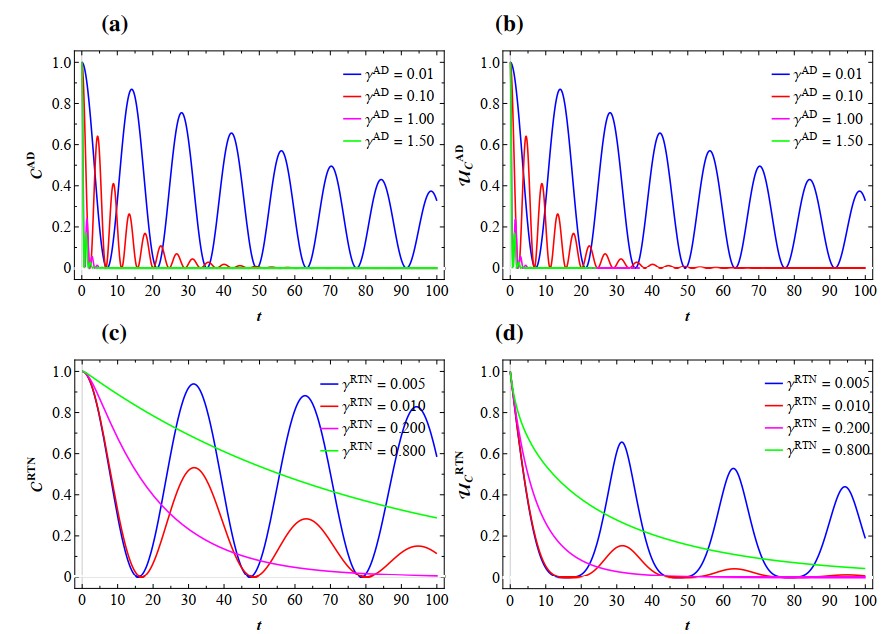} 
		\caption{ The dynamics of (a) concurrence and (b) UIN under non-Markovian AD noise for various values of $\gamma^{\mathrm{AD}}$. The system parameters are: $J_{H}/J_{I} = 1.2$, $\Delta = 0.5$, $k_{B}T/J_{I}=0.01$, $K/J_{I} = 3$, $g = 0.01$, $h_{H}/J_{I} =0$, and $h_{I}/J_{I} = 0$. Panels (c) and (d) present the corresponding dynamics under non-Markovian RTN with $b = 0.05$ respectively.}
		\label{Fig.4}
	\end{figure}
	
	Figure~\ref{Fig.4} provides a systematic investigation into the role of the environmental memory parameter, $\gamma$, in governing the temporal dynamics of quantum correlations within the spin-$\tfrac{1}{2}$ Ising–Heisenberg diamond chain under two distinct local non-Markovian noise processes. The upper panels show the evolution of concurrence, $\mathcal{C}^{\mathrm{AD}}(t,T)$, and UIN, $\mathcal{U}_C^{\mathrm{AD}}(t,T)$, for an AD channel across a range of $\gamma^{\mathrm{AD}}$ values. The lower panels present the corresponding dynamics for $\mathcal{C}^{\mathrm{RTN}}(t,T)$ and $\mathcal{U}_C^{\mathrm{RTN}}(t,T)$ when subjected to a RTN channel with varying $\gamma^{\mathrm{RTN}}$. All simulations are conducted at a fixed cryogenic temperature of $k_B T/J_I = 0.01$ to effectively suppress thermal fluctuations, thereby isolating the influence of environmental memory. For the dissipative AD channel (Figs.~\ref{Fig.4}(a,b)), the persistence of quantum correlations demonstrates a pronounced sensitivity to the value of $\gamma^{\mathrm{AD}}$. In the strongly non-Markovian regime (e.g., $\gamma^{\mathrm{AD}} = 0.01$), both $\mathcal{C}^{\mathrm{AD}}$ and $\mathcal{U}_C^{\mathrm{AD}}$ exhibit sustained, high-amplitude oscillations and unambiguous revivals, which are definitive signatures of significant information backflow from the environment. As $\gamma^{\mathrm{AD}}$ increases, progressing the dynamics toward the Markovian limit, both coherence and entanglement undergo a more rapid decay, and the oscillatory features are systematically attenuated. At high values ($\gamma^{\mathrm{AD}} \geq 1.0$), both metrics display a swift collapse to zero, signifying a transition to a regime dominated by irreversible information loss. This pattern clearly illustrates the preservative capacity of non-Markovian dynamics in mitigating the deleterious effects of dissipation. The dynamics under the pure dephasing RTN channel (Figs.~\ref{Fig.4}(c,d)) exhibit qualitatively similar behavior, yet with notable quantitative differences. For small values of $\gamma^{\mathrm{RTN}}$ (e.g., $0.005, 0.01$), the system once again displays prominent revival phenomena in both concurrence and UIN. However, in contrast to the AD case, these coherent oscillations are more resilient, persisting for longer durations even at intermediate memory parameter values (e.g., $\gamma^{\mathrm{RTN}} = 0.2$), a characteristic that is particularly evident in the concurrence dynamics. As $\gamma^{\mathrm{RTN}}$ becomes large (e.g., $0.8$), the RTN channel enters the Markovian regime, where both quantum correlation measures decay monotonically, with $\mathcal{U}_C^{\mathrm{RTN}}$ degrading more rapidly than $\mathcal{C}^{\mathrm{RTN}}$.
	
	A comparative assessment of the two noise models indicates that the AD channel's dissipative nature induces a more profound degradation of quantum resources, particularly as the Markovian limit is approached. In contrast, the RTN channel, being a population-conserving dephasing mechanism, is more conducive to the preservation of coherence over extended periods. Across both noise paradigms, the concurrence, $\mathcal{C}$, consistently exhibits more pronounced and earlier revival features than UIN, suggesting that $\mathcal{C}$ functions as a more sensitive probe for high-frequency coherence dynamics, whereas UIN reflects broader, more persistent non-classical correlations. Our analysis offers a complementary perspective to recent work by Carrion et al.~\cite{R2025}, which focused on different metrics under Markovian decoherence at zero temperature. By contrast, the present study emphasizes entanglement and UIN within a distinct spin architecture under non-Markovian and finite-temperature conditions, thereby broadening the understanding of quantum correlation dynamics in realistic settings.
	
	 Figure~\ref{Fig.4} establishes that the lifetime and dynamical structure of quantum correlations are critically determined by the environmental memory time. A stronger non-Markovian character (manifested as smaller $\gamma$) significantly enhances the persistence of these correlations by enabling mechanisms such as information backflow and coherence revivals. These results underscore the pivotal importance of reservoir engineering in the development of quantum technologies, proposing that the strategic control over system-environment memory parameters constitutes a pragmatic strategy for extending the functional viability of quantum information devices.
	
	\section{Conclusion and Outlook}
	\label{Sec6}
	
This paper provided a comprehensive analysis of quantum information dynamics within a spin-one-half Ising-Heisenberg diamond chain subjected to local environmental interactions characterized by memory effects. The study considered both dissipative amplitude damping and pure dephasing random telegraph noise channels. By exploring the behavior of two fundamental quantum resources—entanglement, as quantified by concurrence, and a broader measure of non-classicality, captured by uncertainty-induced nonlocality—this work revealed how these correlations evolved under the influence of thermal fluctuations, internal spin couplings, external magnetic fields, and the specific character of environmental memory.
	
Our results demonstrate that the nature of the decoherence channel critically determined the preservation and revival of quantum correlations. Specifically, pure dephasing noise proved to be less destructive to quantum coherence than dissipative noise, which permitted smoother and more enduring revival cycles for both entanglement and the uncertainty-based nonlocality metric. Furthermore, non-Markovian memory effects exerted a protective influence under both types of noise and facilitated a temporary retrodiction of information from the environment that transiently restored coherence and entanglement. Notably, the internal Heisenberg exchange coupling and the four-spin interaction term significantly bolstered the robustness of quantum correlations, which highlighted their utility as intrinsic control parameters for enhancing coherence.
	
Temperature emerged as a decisive factor in the degradation of quantum correlations. While both entanglement and uncertainty-based nonlocality decayed with increasing thermal energy, the latter metric exhibited superior persistence at elevated temperatures, even in regimes where entanglement was entirely suppressed. This observation underscored the broader applicability of uncertainty-induced nonlocality as a diagnostic tool for detecting non-classical behavior in thermally active and noisy systems. Additionally, the application of external longitudinal magnetic fields introduced a non-monotonic thermal behavior, whereby moderate thermal energies could optimize quantum correlations due to field-induced level reconfigurations.
	
	From the perspective of quantum information technologies, our findings carry several important implications. First, the observed resilience and revivals of quantum correlations under non Markovian noise support the design of robust quantum memories, communication protocols, and entangled state preparation schemes in solid state spin networks. Second, the capacity of UIN to detect quantum correlations in high temperature regimes and beyond entanglement points to its potential use in quantum metrology, where sensitivity to coherence is critical. Third, our results highlight the strategic role of engineered couplings and magnetic field control in tailoring quantum coherence in hybrid systems. Importantly, the hybrid Ising, Heisenberg diamond model considered here can be emulated in systems such as superconducting qubits with engineered couplings, Rydberg atom arrays, or trapped ions, where both Ising and Heisenberg interactions have been demonstrated. Our results may thus inform the design of experiments probing decoherence and quantum memory resilience in noisy quantum processors.
	
	Furthermore, the temperature threshold at which all initial entanglement was thermally suppressed could be interpreted as a crossover point, marking a transition from a quantum-coherent regime to a classically dominated one. Although not a phase transition in the strict thermodynamic sense, this threshold signified a qualitative shift in the system's underlying correlation structure. It was suggested that future investigations could explore this connection further by linking this characteristic temperature to measurable thermodynamic observables, such as specific heat or magnetic susceptibility.
	
	The study concluded by identifying several promising research trajectories. A natural extension was proposed to involve the study of multipartite correlations and quantum steering within larger spin clusters. Another suggested direction concerned the incorporation of time-dependent control fields or advanced noise-engineering techniques to dynamically manage entanglement lifetimes. Further questions were identified for future work, including how multipartite entanglement behaves in this system and whether time-varying couplings can enhance resilience against decoherence. Finally, it was concluded that experimental realizations using quantum simulators, including trapped ions, superconducting circuits, or solid-state defects, would provide a crucial validation of these theoretical predictions and guide the development of next-generation, noise-resilient quantum architectures.

	\section*{Appendix A}
	\label{Sec7}
	In the Fano-Bloch representation, the density matrix $\rho^{X}_{\alpha}(t,T)$, corresponding to either the amplitude damping (AD) channel (\ref{R1}) or the random telegraph noise (RTN) channel (\ref{R2}), can be expressed as follows:
	
	\begin{equation}
		\rho^{X}_{\alpha}(t,T)=\frac{1}{{4}}\sum\limits_{n ,m =0}^{{3}}\mathcal{R}^{X}{{{_{n, m
					}(t,T)}}}\left( {\sigma}^{n}{{\otimes {\sigma^{m}}}}\right) ,
	\end{equation}%
	where $\mathcal{R}^{X}{{{_{n, m }(t,T)}}}=\mathrm{Tr}\left( \rho^{X}_{\alpha} (t,T)({\sigma }%
	^{n }{{\otimes {\sigma ^{m })}}}\right) $ are the total correlation
	tensor components occurring in the Fano-Bloch decomposition associated with
	bipartite density matrix $\rho^{X}_{\alpha}(t,T)$ with ${\sigma }^{n }({{{\sigma ^{m }}%
	}})=\{{\sigma }^{0},{\sigma }^{i }({\sigma }^{j})\}.$ The
	non-vanishing components $\mathcal{R}^{X}{{{_{n, m }(t,T)}}}$ are calculated as 
	\begin{eqnarray}
		\mathcal{R}^{X}_{0, 0} &=&\mathrm{Tr}\text{ }\rho^{X}_{\alpha}(t,T){=1},\text{ \ }\mathcal{R}^{X}%
		_{0, 3} =1-2{}(\rho^{X}_{2,2}+\rho^{X}_{4,4}),\text{ \ \ }\mathcal{R}^{X}_{1,1}=%
		\mathcal{R}^{X}_{2,2}=|\rho^{X}_{2,3}|,\text{ \ }  \notag \\
		\mathcal{R}^{X}_{3,0}&=&1-2(\rho^{X}_{3,3}+\rho^{X}_{4, 4}),\text{ \ }\mathcal{R}^{X}%
		_{3, 3}= 1-2(\rho^{X}_{2,2}+\rho^{X}_{3,3}).
	\end{eqnarray}%
	\ Therefore, in terms of the Fano-Bloch components $\mathcal{R}^{X}{{{_{n m
				}(t,T)}}}$ associated with the matrix $\rho^{X}_{\alpha} (t,T)$ $(\ref{R1}-\ref{R2})$, the
	eigenvalues of the $3\times 3$ matrix $(W^{X}_{AB})_{ij} (t,T)$ \ Eq. ($\ref{W}$) can be expressed as
	
	\begin{eqnarray}
		\mathcal{W}^{X}_{11}(t,T) &=&\mathcal{W}^{X}_{22}(t,T)=\sqrt{\left( \Xi^{X}_{+}+2\sqrt{d^{X}_{1}}\right) \left( \Xi^{X}_{-}+2\sqrt{d^{X}_{2}}\right) }+\frac{1}{4}\frac{%
			\left( \mathcal{R^2}^{X}_{03}-\mathcal{R}^{X^2}_{30}\right) }{\sqrt{\left(
				\Xi^{X}_{+}+2\sqrt{d^{X}_{1}}\right) \left( \Xi^{X}_{-}+2\sqrt{d^{X}_{2}}\right) }},
		\notag \\
		\mathcal{W}^{X}_{33}&=&\frac{1}{2}\left( 1+2\left( \sqrt{d^{X}_{1}}+\sqrt{%
			d^{X}_{2}}\right) \right) + \\
		&&\frac{1}{8}\left( \frac{\left( \mathcal{R}^{X}_{03}+\mathcal{R}^{X}%
			_{30}\right) ^{2}-\left( \mathcal{R}^{X}_{11}-\mathcal{R}^{X}_{22}\right)
			^{2}}{\left( \Xi^{X}_{+}+2\sqrt{d^{X}_{1}}\right) }+\frac{\left( \mathcal{R}%
			^{X}_{03}-\mathcal{R}^{X}_{30}\right)^{2}-\left( \mathcal{R}^{X}_{11}+\mathcal{R}^{X}_{22}\right) ^{2}}{\left( \Xi^{X}_{-}+2\sqrt{d^{X}_{2}}\right) }\right),
	\end{eqnarray}
	where 
	\begin{eqnarray}
		\Xi^{X}_{\pm} &=&\frac{1}{2}\left( \mathcal{R}^{X}_{00}\pm \mathcal{R}^{X}_{03}\right) , \\
		d^{X}_{1,2} &=&\frac{1}{16}\left[ \left( \mathcal{R}^{X}_{00}\pm \mathcal{R}^{X}%
		_{33}\right) ^{2}-\left( \mathcal{R}^{X}_{30}\pm \mathcal{R}^{X}_{03}\right) ^{2}-\left( \mathcal{R}^{X}_{11}\mp \mathcal{R}^{X}_{22}\right) ^{2}\right].  \notag
	\end{eqnarray}
	Hence, in this case, it is easy to check that the matrix $(W^{X}_{AB})_{ij}(t,T)$ \ ($\ref{W}$) is diagonal ($\mathcal{W}^{X}_{12}(t,T)=%
	\mathcal{W}^{X}_{21}(t,T)=\mathcal{W}^{X}_{23}(t,T)=\mathcal{W}^{X}_{32}(t,T)=\mathcal{W}^{X}_{13}(t,T)=\mathcal{W}^{X}_{31}(t,T)=0$). Thus, its eigenvalues are simply reduced
	to diagonal elements $(W^{X}_{AB})_{ii}(t,T)$ ($i=1,2,3$). 
	
\section*{Data availability statement}
No data available in this study.

\end{document}